\documentclass[sigconf]{acmart}
\usepackage[hide]{chato-notes} %  change show -> hide
\usepackage{subcaption}
\usepackage{graphicx}
\usepackage{comment}

\copyrightyear{2021}
\acmYear{2021}
\setcopyright{rightsretained}\acmConference[To appear in CIKM'21]{}{}{}
\acmBooktitle{}
\acmPrice{}

\begin{document}

\title{WikiCheck: An end-to-end open source Automatic Fact-Checking API based on Wikipedia}


\author{Mykola Trokhymovych}
\affiliation{%
  \institution{Ukrainian Catholic University}
%  \streetaddress{1 Th{\o}rv{\"a}ld Circle}
  %\city{ekla}
  \country{Ukraine}}
\email{trokhymovych@ucu.edu.ua}

\author{Diego Saez-Trumper}
\affiliation{%
  \institution{Wikimedia Foundation}
  %\city{Rocquencourt}
  \country{Spain}}
\email{diego@wikimedia.org}

%renewcommand{\shortauthors}{Trovato and Tobin, et al.}

\begin{abstract}
 %\sloppy 
 %The incoming flow of information is continuously increasing along with the disinformation piece that can harm society.
 %Filtering unreliable content helps keep Wikipedia as free as possible of disinformation, making it one of the most significant reliable information sources. Consequently, Wikipedia's knowledge base is widely used for facts verification academic research. The main goal of our work is to transform recent academic achievements into a practical open-source Wikipedia-based fact-checking application that is both accurate and efficient.  We observe the primary NLI related datasets and study their relevant limitations. As a result, we propose the data filtering method that improves the model's performance and generalization. We show that transfer learning for NLI models are not working well, and complete model training is needed to achieve the best result on a specific dataset. We come up with an unsupervised fine-tuning of the Masked Language model on field-specific texts for model domain adaptation. Finally, we present the new fact-checking system $WikiCheck$ API that automatically performs a facts validation process based on the Wikipedia knowledge base. It is comparable to SOTA solutions in terms of accuracy and can be used on low memory CPU instances.%, making it usable in production environments. % that make it applied to the field of fact-checking.
With the growth of fake news and disinformation, the NLP community has been working to assist humans in fact-checking. However, most academic research has focused on model accuracy without paying attention to resource efficiency, which is crucial in real-life scenarios. In this work, we review the State-of-the-Art datasets and solutions for Automatic Fact-checking and test their applicability in production environments. We discover overfitting issues in those models, and we propose a data filtering method that improves the model's performance and generalization. Then, we design an unsupervised fine-tuning of the Masked Language models to improve its accuracy working with Wikipedia. We also propose a novel query enhancing method to improve evidence discovery using the Wikipedia Search API. Finally, we present a new fact-checking system, the \textit{WikiCheck} API that automatically performs a facts validation process based on the Wikipedia knowledge base. It is comparable to SOTA solutions in terms of accuracy and can be used on low-memory CPU instances.
 
\end{abstract}

\begin{CCSXML}
<ccs2012>
   <concept>
       <concept_id>10002951.10003317</concept_id>
       <concept_desc>Information systems~Information retrieval</concept_desc>
       <concept_significance>500</concept_significance>
       </concept>
 </ccs2012>
\end{CCSXML}

\ccsdesc[500]{Information systems~Information retrieval}

\keywords{Wikipedia, fact-checking, NLI, NLP, applied research}


\maketitle

\section{Introduction}
    
Disinformation, fake news, and weaponization of information have become well-known concepts in recent years~\cite{singer2018likewar,o2016weapons}.  Computer scientists (CS), and more specifically the Natural Language Processing (NLP) community are trying to contribute to the fight against disinformation by creating Automated Fact-Checking Systems (AFCS) that can assist humans in the process of validating or rejecting a piece of information~\cite{claimbooster}. 

At the same time, beyond computer scientists, other communities have been working in creating trustworthy sources of information. While the idea of "absolute truth" is debatable, communities like Wikipedia editors have emphasized verifiability and neutrality of information~\cite{saez2019online}. Understanding the source and the peer-validation of information are key concepts to navigate knowledge. As using traceable information, coming from reliable sources is the way that \textit{Wikipedians} had found to build one the most extensive knowledge bases in the world, one could expect that AFCS do the same. Therefore, transparency, and explainability of ML-based solutions, are essential to develop a trustworthy AFCS~\cite{smith2020keeping,halfaker2020ores}.

However, nowadays, despite the large efforts done by the NLP researchers, it is challenging to find  ML packages or systems that can perform the complete task of fact-checking, from receiving an open-domain claim and classifying it as true or false. First of all, just a few datasets can be used to train and validate such systems. Some of the most popular datasets are entirely synthetic and are difficult to extrapolate to real-life scenarios.   Moreover, existing datasets contain data artifacts that create problems with model generalization~\cite{ref_lncs6}. Additionally, most models developed on academic research focus just on accuracy, underestimating problems such as time and resource constrains. 

In this work,  we tackle the problem of creating an end-to-end open-source system for Automated Fact-Checking that can be used on production environments, the \textit{WikiCheck} API. First of all, we review existing State-of-the-Art (SOTA) solutions, testing their applicability to real scenarios. We confirm known issues and uncover new artifacts on the currently most used datasets. Utilizing Transfer learning and designing efficient heuristics, we propose a methodology to use existing datasets to train models that can be used in real-life scenarios. Next, we focus on openness and reproducibility and build a fact-checking system on top of open-source libraries, and rely on open knowledge sources such as Wikipedia. Our system deals with the trade-off between efficiency and accuracy, working in small CPU virtual machines with a SOTA-comparable performance.  We release our solution in the form of an open API\footnote{ \url{https://nli.wmflabs.org}}, and also share all the code to make it easy to reproduce our system and results\footnote{\url{https://github.com/trokhymovych/WikiCheck}}.

\begin{figure*}[!t]
\centering
\includegraphics[width=\textwidth]{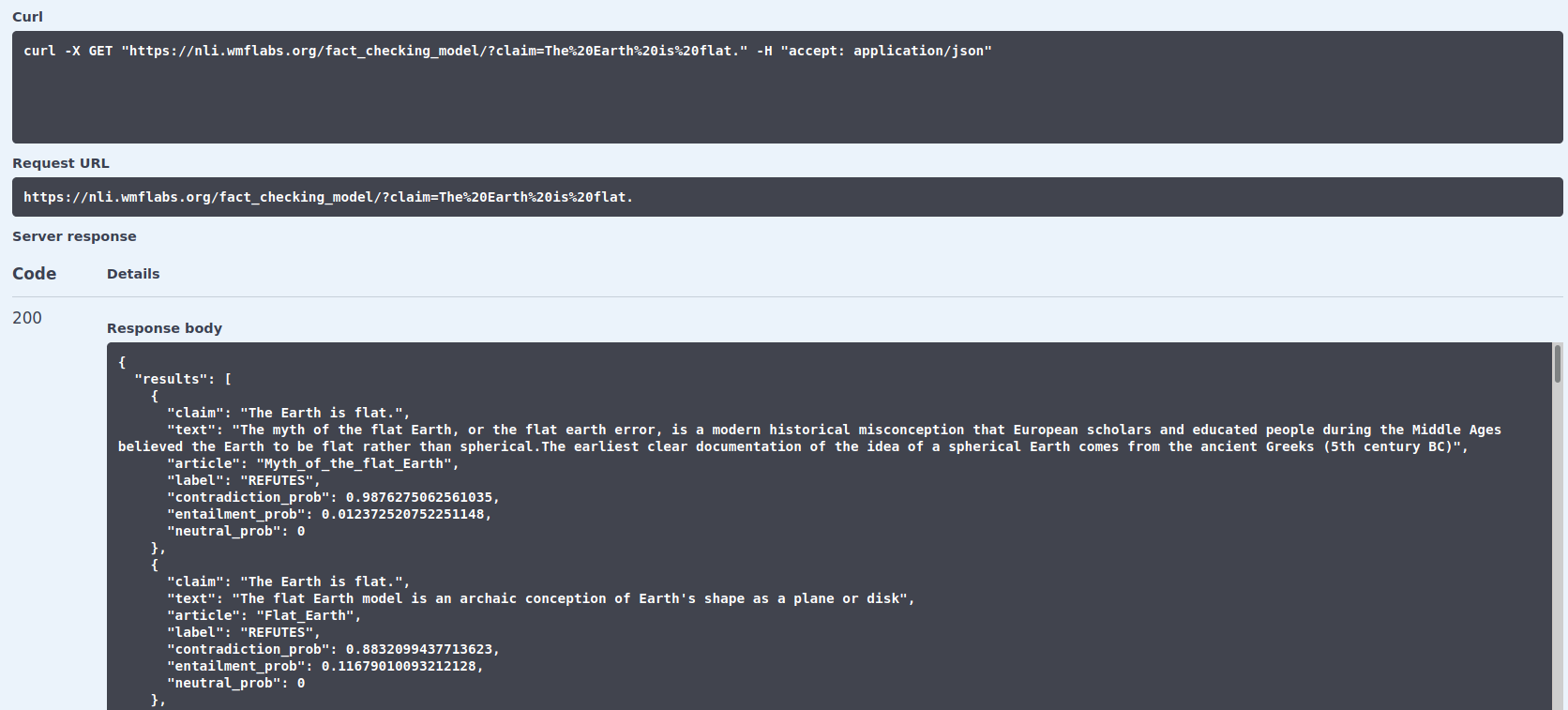}

\caption{Live example of an API call on the WikiCheck Instance running in \url{https://nli.wmflabs.org}.}
\vspace{-12pt}
\label{fig:screenshot}
\end{figure*}

\section{Related work}
\label{related:problem}

The task of fact-checking was initially used in journalism as an essential part of news reporting.  In 2014, one of the first datasets published on this domain consisted of 221 labeled claims - related to politics - manually checked by Politifact\footnote{\url{https://www.politifact.com/}} and Channel4 with related sources of evidence~\cite{ref_lncs24}. After that, in 2017, similar data collection but much more extensive, containing 12.8K labeled claims from Politifact, was released by~\cite{ref_lncs25}.%However, this can be considered a small collection of data to train large deep-learning models.

For computer scientists, the task of comparing two pieces of text and deciding whether one supports or rejects the other is known as Natural Language Inference (NLI)~\cite{bowman2015large}. However, the datasets mentioned above are small and complex to train deep-learning models. For that reason,  NLI-focused  large-datasets has been created, such as the benchmark ones SNLI~\cite{ref_lncs5} and MNLI~\cite{ref_lncs26}, where a pair of claims are labeled with  \textit{entailment}, \textit{contradiction}, or \textit{neutral}. Alternatively, efforts like the  WIKIFACTCHECK-ENGLISH~\cite{ref_lncs12} and FEVER~\cite{ref_lncs7} datasets tried to give a more realistic approach, using Wikipedia as part of their claim or evidence creation. In Sec. \ref{chap3} we discuss these three datasets in detail.

\subsection{Masked language modeling}
The most crucial part of NLI solutions is language models. The recent SOTA solutions are built on top of them. 

One of the most valued recent contributions to NLP is the BERT architecture \cite{ref_lncs8}. 
The BERT model made a revolution in the NLP field. It significantly moved SOTA scores for several NLP tasks by presenting new architecture for language modeling. It showed point absolute improvement 7.7\% on GLUE score \cite{ref_lncs11}. The authors present a solution allowing it to be bidirectional and utilize the masked language model (MLM) and next sentence prediction (NSP) as a pretraining objective. The MLM training process is built on masking some of the tokens and predicting them based only on their context \cite{ref_lncs8}. Training using NSP loss is working by choosing the two sentences as training sample, 50\% of the time, one is following another one (labeled as IsNext), then another half represented by random sentences from the corpus (labeled as NotNext). Training models with NSP loss is beneficial for NLI problem according to~\cite{ref_lncs8}. 

Another relevant work is  Sentence-BERT~\cite{ref_lncs10}. Authors present a way to train sentence embeddings instead of word embeddings, improving the model's efficiency, and making transformer models like BERT possible to use in high-load production tasks.

\subsection{State of the art solution} \label{sota_solution}
In this subsection, we observe works in the NLI field.
We divided SOTA solutions in two groups: \emph{(i)} sentence-based and \emph{(ii)} word-based.
Sentence-based solutions are usually faster, more applicable in real life as vectors can be cached. However, word-based solutions are more precise according to SNLI published results comparison\footnote{\url{https://nlp.stanford.edu/projects/snli/}}. 

The most recent word-based solutions use the MLM, additional features, and various training strategies to achieve high accuracy results. 
Multi-Task Deep Neural Networks for NLU propose training BERT model on multiple NLU tasks simultaneously, benefiting from a regularization and achieving 91.6\% on SNLI~\cite{ref_lncs16}. 
SemBERT presents integrating contextualized features into language model, extending it with semantics~\cite{ref_lncs19}. It is also fine-tuned separately for different tasks achieving 91.9\% on SNLI.
CA-MTL demonstrates faster fine-tuning as most parameters are frozen and the dataset balanced across different tasks, reaching the current NLI SOTA of 92.1\% on SNLI~\cite{ref_lncs20}.

The top-performing sentence-based solutions are older and less accurate. Also, they are not using MLM comparing with word-based SOTA. Sentence Embeddings in NLI with Iterative Refinement Encoders achieve 86.6\% on SNLI using hierarchical BiLSTM model with Max Pooling and iterative refinement~\cite{ref_lncs18}. One more approach is using dynamic meta-embeddings out of Word2Vec \cite{ref_lncs21} or fasttext \cite{ref_lncs22} for sentence embeddings composition, reaching 86.7\% \cite{ref_lncs17}. Current NLI sentence-based SOTA is 88.6\%, achieved using BiLSTM block to represent words and context~\cite{ref_lncs13}.
We see that scores for those types of models are lower than word-based. However, such models can be used in production high-load tasks as they are lighter and faster. Also, there are achievements in the word-based approach that can be transferred to sentence-based models to improve them.

\subsection{End-to-end fact verification solutions}

One effort to create an end-to-end system is Claimbuster \cite{claimbooster}. However, that solution relies on external - closed - components and is not tested against benchmark datasets. More recently, a pre-print shared by \cite{chernyavskiy2021whatthewikifact} proposed a high-level solution considering a more complete conceptual pipeline.  In contrast, in this work, we develop and deploy an end-to-end system that can be used as an open API and that is tested on benchmark datasets \cite{fever_baseline,ref_lncs26,bowman2015large}.

A relevant bulk of work creating stand-alone systems is the solutions designed for the FEVER Shared task ~\cite{fever_task}. The FEVER challenge was to implement an automated fact verification system. It differs from standard NLI formulation, as here we need not just classify relation between two sentences but also pick the evidence (hypothesis) sentence from a knowledge base. That makes this task more complicated and closer to the real-world scenario at the same time. %We will analyze the top solutions of that task in this section. 
As a baseline system presented by \cite{fever_baseline}, most of the solutions are multistage models that perform document retrieval, sentence selection, and sentence classification. Baseline exploits TF-IDF-based retrieval to find the relevant evidence and an NLI model to classify the relationship between the returned evidence and the claim.

The UNC-NLP solution achieves 0.64 FEVER score comparing to 0.28 of baseline \cite{fever_baseline,fever1}. It uses neural models to perform deep semantic matching for both document and sentence retrieval stages and adds WordNet features to improve the NLI model. UCL MRG team proposes to use logistic regression for the document and sentences retrieval stage and treat the aggregation phase as an additional stage~\cite{fever2}. Athene team achieves a 0.61 FEVER score with entity linking and WikiMedia search API for article search and Glove and FastText embeddings for the NLI model~\cite{fever3}. In Sec.~\ref{chap5} we compare our results against these systems.

\section{Data exploration}
\label{chap3}

% \begin{figure}[!t]
% \centering
% \includegraphics[width=.23\textwidth]{pictures/i8.png}\hfill
% \includegraphics[width=.23\textwidth]{pictures/i9.png}\hfill
% \caption{Distribution of length of hypothesis in training dataset of SNLI (top) and MNLI (bottom)}
% \vspace{-12pt}
% \label{fig:figure3}
% \end{figure}

\begin{figure}[!t]
\centering
\includegraphics[width=.49\textwidth]{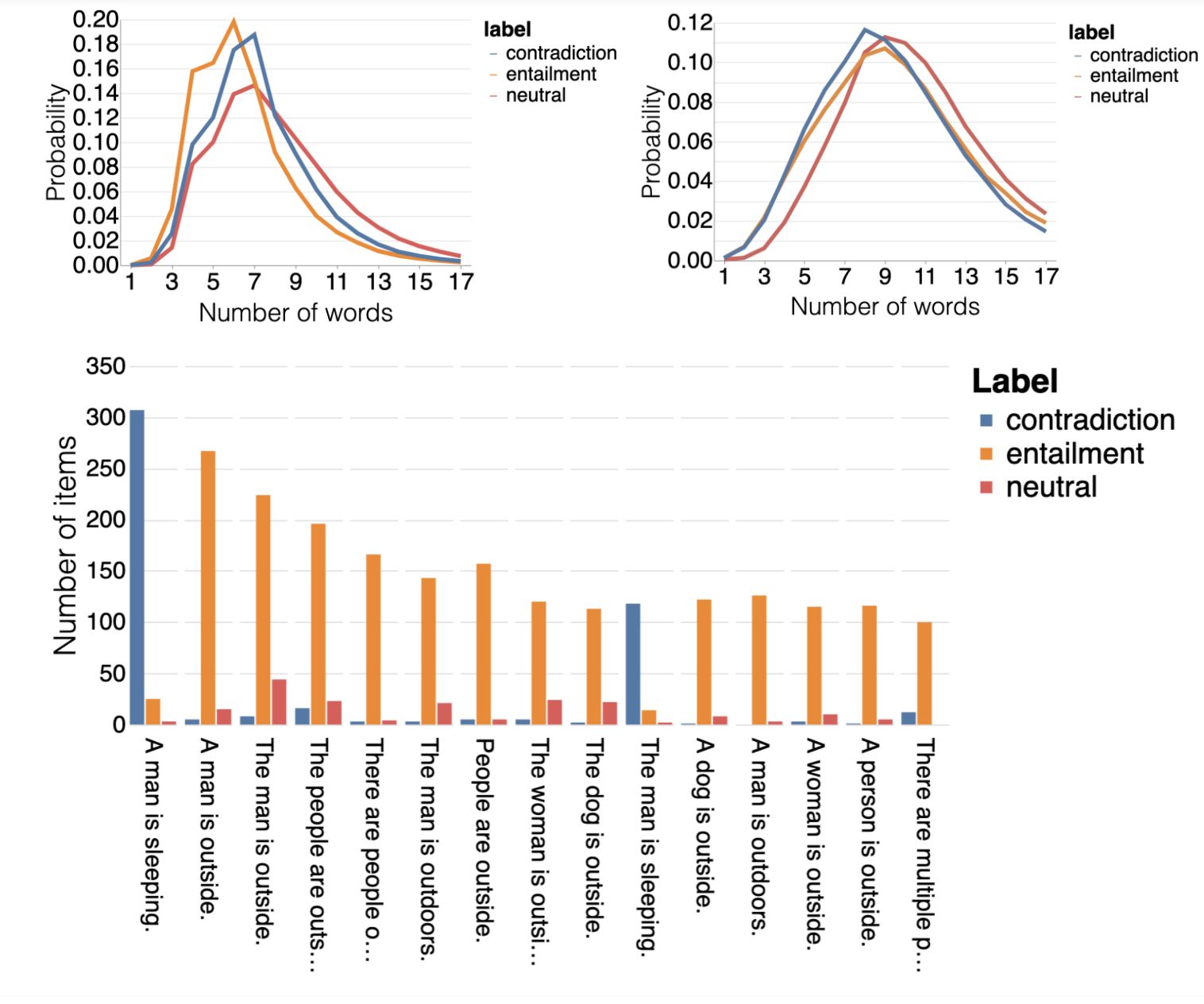}\hfill
\caption{Distribution of length of hypothesis in training dataset of SNLI (top right) and MNLI (top left), SNLI dataset top-15 the most frequent hypothesis and their classes counts (bottom)}
\vspace{-12pt}
\label{fig:figure3}
\end{figure}

The most recent research in the NLP field is heavily dependent on data. Here, we observe and discuss the primary datasets used to solve the NLI problem and their characteristics. 

\subsection{General purposes datasets: SNLI and MNLI}
SNLI and MNLI are the most used benchmark datasets. They allow us to compare results with the most recent SOTA solutions. Both datasets consist of a claim, a related hypothesis, and a label, either \textit{neutral}, \textit{contradiction}, or \textit{entailment}.

SNLI dataset comes from image captions when MNLI is from a wide range of styles, degrees of formality, and topics~\cite{ref_lncs26}.

\begin{comment}

\begin{table}[t]
\begin{center}
\caption{Samples from MNLI and SNLI datasets}
{\tabcolsep=3pt
\label{table4}
\begin{tabular}{|p{1cm}|p{2.9cm}|p{1.9cm}|p{1.7cm}|}
\hline
\textbf{Dataset} & \textbf{Claim} & \textbf{Hypothesis} & \textbf{Label}\\
\hline
MNLI & The Old One always comforted Ca'daan, except today. & Ca'daan knew the Old One very well. & \textit{neutral}\\
\hline
MNLI & 
At the other end of Pennsylvania Avenue, people began to line up for a White House tour. & People formed a line at the end of Pennsylvania Avenue. & \textit{entailment}\\
\hline
SNLI & A man inspects the uniform of a figure in some East Asian country. & The man is sleeping & \textit{contradiction} \\
\hline
SNLI & An older and younger man smiling. & Two men are smiling and laughing at the cats playing on the floor. & \textit{neutral} \\
\hline
\end{tabular}%
}
\end{center}
\end{table}
\end{comment}

It is important to mention that all classes are well-balanced.  We analyzed distributions of the length of three classes' claims and hypotheses and found that the claims' length is equally distributed. At the same time lengths of the hypotheses are different within different classes. Figure \ref{fig:figure3} shows that the \textit{entailment} class hypothesis is usually shorter than others. It can influence the model that could learn the length of a sentence instead of its meaning. In MNLI, distributions of the lengths are more balanced, but the \textit{neutral} class sentences are usually longer. Moreover, we see that the MNLI hypotheses are larger than SNLI.

In our exploration, we found out the reason why certain words in the hypothesis are highly correlated with specific classes, as was discussed by~\cite{ref_lncs6}. We defined top-15 the most frequent hypothesis used by annotators and analyzed the classes to which they correspond. We found out that frequent hypotheses are usually used in either \textit{entailment} or \textit{contradiction} class, represented in Figure~\ref{fig:figure3}.  It is also not natural behavior as the model will learn only the sense of hypothesis instead of the desired relation between claim and hypothesis. In Sec. \ref{chap5} we analyze how to filter out such patterns. %from training will 

% \begin{figure}[!t]
% \centering
% \includegraphics[width=0.48\textwidth]{pictures/i4.png}\hfill
% \caption{SNLI dataset top-15 the most frequent hypothesis and their classes counts}
% \label{fig:figure4}
% \vspace{-12pt}
% \end{figure}

\subsection{Wikipedia specific dataset: FEVER}
\sloppy
There are two main Wikipedia-related datasets for NLI: WIKIFACTCHECK \cite{ref_lncs12} and FEVER \cite{fever_task}. In this work, we focus on the latter because its best tries to reproduce a real-life scenario. Also, we explored WIKIFACTCHECK finding several annotation artifacts that we omit here due to lack of space.

FEVER dataset consists of 185,445 claims generated by altering sentences extracted from Wikipedia and subsequently verifying without knowing the sentence they were derived from. The claims are classified as \textit{SUPPORTS}, \textit{REFUTES}, or \textit{NOT-ENOUGH-INFO} by annotators \cite{ref_lncs7}. Hypotheses supporting the claims are sentences from the summary section of related articles. This dataset differs from previous ones, as it represents another problem formulation because it not only implies classifying the relation between two pieces of text but also linking the given claim with corresponding evidence in a  knowledge base. That knowledge base is a Wikipedia dump.  FEVER task is a more generalized problem setting closer to the real-life scenario, simulating what humans could do to fact-check a given claim. This dataset is the main validation for our research.

The original dataset consists of a claim, label, and evidence link. In case the label is \textit{NOT ENOUGH INFO}, there is no corresponding link. The evidence link is the name of the Wikipedia article with the number of the sentence in that article. Also, we have the actual Wikipedia dump that allows us to find out the evidence sentence. In our case, we used the FEVER dataset to build two types of datasets:  The first one consists of a claim and article link as in Table~\ref{table_fever1}. The second one contains a claim, corresponding evidence sentence from Wikipedia dump and label. That is the SNLI style of the dataset with only one difference that we have two classes: \textit{REFUTES} and \textit{SUPPORTS}— a sample of such data presented in Table~\ref{table_fever2}.

%%%%%%%%%%%%%%%%%%%%%%
\begin{table}
\begin{center}
\caption{FEVER data sample. Article linking.}
{\tabcolsep=3pt
\begin{tabular}{|p{3.5cm}|p{4.5cm}|} %14.5
\hline\label{table_fever1}
\textbf{Claim} & \textbf{Evidence Articles}\\
\hline
Nikolaj Coster-Waldau worked with the Fox Broadcasting Company. & Fox\_Broadcasting\_Company, Nikolaj\_Coster-Waldau\\
\hline
%Hermit crabs are arachnids. & Arachnid, Hermit\_crab, Decapoda\\
%&\hline
There is a capital called Mogadishu. & Mogadishu\\
\hline
\end{tabular}%
}
\end{center}
\end{table}

%%%%%%%%%%%%%%%%%%%%%%%%%%%%%%%
\begin{table}
\begin{center}
\caption{FEVER data sample. SNLI-style relation dataset.}
{\tabcolsep=3pt
\begin{tabular}{|p{1.3cm}|p{5.1cm}|p{1.4cm}|}
\hline\label{table_fever2}
\textbf{Claim} & \textbf{Hypothesis} & \textbf{Label}\\
\hline
Roman Atwood is a content creator.&He is best known for his vlogs, where he posts updates about his life daily.&SUPPORTS\\
\hline
Adrienne Bailon is an accountant.&Adrienne Eliza Houghton (née Bailon; born October 24, 1983) is an American singer-songwriter, recording artist, actress, dancer, and television personality... &REFUTES\\
\hline
\end{tabular}%
}
\end{center}
\end{table}

\section{\textit{WikiCheck} architecture}

Here, we propose our fact-checking system architecture. We decompose the application into two major parts: the candidate's selection model (level one) and the NLI classification model (level two).

\begin{figure*}[!t]
\centering
\includegraphics[width=0.97\linewidth]{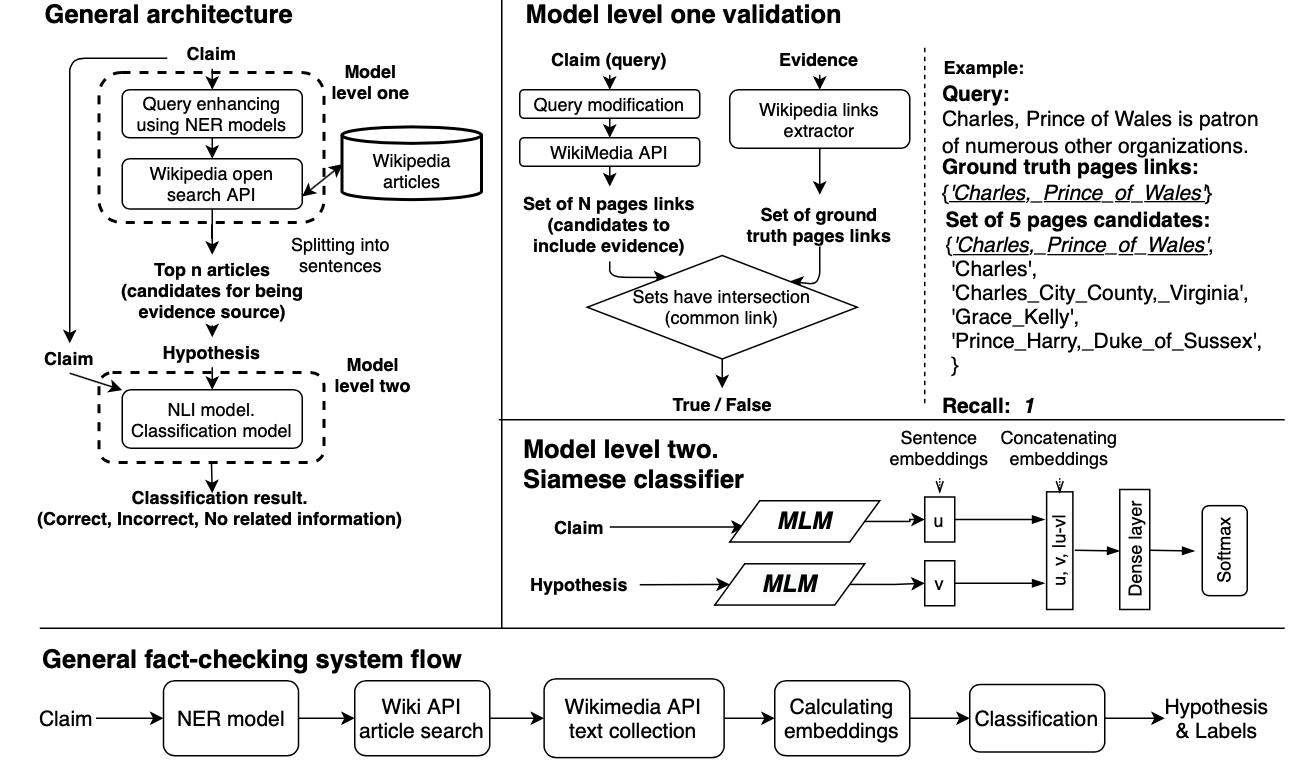}\hfill
\caption{Full overview of the WikiCheck system architecture.}
\label{fig:figure_solution}
\end{figure*}

The main idea of our approach is to reproduce the human way to do the fact-checking process: the input is a fact in the form of text (claim), next, we search for evidence (hypothesis) from a trusted source and comparing the claim, and the evidence concludes that we can support, reject or if that there is not enough evidence to decide. Based on this procedure, we propose our architecture (Fig~\ref{fig:figure_solution}).

We receive a claim as input and use the MediaWiki API\footnote{ \url{https://www.mediawiki.org/wiki/API:Search}} to extract related Wikipedia articles.
Here we experiment with query enhancing techniques to improve the search recall. Next, we split each of the retrieved articles into sentences and pass to the second stage model. We compare the claim and the set of hypothesis sentences using an NLI model to define if the input claim is correct, incorrect, or we have no related information in our knowledge base about it. 

\subsection{Model level one. Wikipedia search API}

The first level model performs a full-text search through the whole (English) Wikipedia.
Although we cannot influence the search engine, which is a significant limitation of our approach, we can improve the query itself. We do this using a NER approach to enrich the query. In section \ref{chap5} we discuss the techniques used on query optimization.

At this stage, we use the \textit{SUPPORTS}, and \textit{REFUTES} classes on the FEVER dataset to validate our NER-based solution. Given that FEVER was created in 2017 and that we are using the live MediaWiki API (with current Wikipedia content), we needed to filter out around 4.3\% of the claims because the evidence articles were unreachable at the time of our experiments (2020). 

The validation process itself is represented in Figure \ref{fig:figure_solution} with corresponding example. The general idea is given a claim, we pick up a set of candidate articles using MediaWiki API. Then we compare the obtained set with ground truth items provided in the FEVER dataset. 
We use the Average Recall (AR) metric for results validation and comparison. 

The AR is just the average of recall metrics for each search. This metric does not consider the negative samples retrieved, but we are interested in all relevant candidates finding.

\subsection{Model level two. Natural language inference model} \label{model_two}

Model level two is where we apply our NLI model with  three possible outputs: \textit{SUPPORTS}, \textit{REFUTES}, and \textit{NEI}. 

The general idea of the presented NLI model is a Siamese network using a BERT-like model as a trainable encoder for sentences. The idea comes from \cite{ref_lncs27}, with the difference that we are not using multiplication of sentence vectors in concatenation layer, but using the original vectors and their absolute difference following \cite{ref_lncs10}'s approach. 

In our problem formulation, we need to compare one claim with multiple hypothesis sentences. The sentence-based model approach enables us to calculate claim embedding only once and reuses it for every hypothesis. Also, embeddings for the hypothesis can be batch processed and precalculated in advance.  
The presented architecture enables to cache intermediate results of embeddings and reuse them for online prediction. In the following sections, we present the results of our experiments with different BERT-like encoders and training strategies.

\section{Experiments and validation}
\label{chap5}

\subsection{Improving the performance of search}

At this level, our main challenge is to find relevant Wikipedia articles regarding a given claim. 
The excellent articles retrieval system return the minimum number of results, with the highest Average Recall (AR). The number of returned results directly influences the efficiency of the next stage, as the more texts we retrieve, the more time we need to encode them using an MLM encoder. 

Moreover, increasing the number of candidates (larger $N$) does not improve recall significantly.  We tested with $\approx100K$ queries, obtaining, on average, an $AR = 0.628$ for $N=10$. We tried with $N$ equal to 30 and 50, with the marginal improvement of 2\% in AR (Table \ref{table5}).

Another way to improve the recall was to enhance the query. We tested a set of NER packages and two strategies to include the recognized entities to our search: \textit{(i)} merge all named entities in one query; or \textit{(ii)} create separated queries for each entity we recognized. Specifically, we used the \textit{"en\_core\_web\_sm"} (\textit{"sm"}) and \textit{"en\_core\_web\_trf"}~(\textit{"trf"}) from spaCy\footnote{\url{https://spacy.io}} and \textit{"ner-fast"} model from Flair\footnote{\url{https://github.com/flairNLP/flair}}.

\begin{table}[t]
\begin{center}
\caption{Comparing Level one configurations performance for different $N$, NER systems and aggregation strategies. }
\label{table5}
{\tabcolsep=3pt
\begin{tabular}{p{4cm}|p{1.6cm}|p{1.8cm}|}
\hline
\textbf{Configuration} & \textbf{AR (higher is better)}  &  \textbf{N returned, (lower is better)}\\
\hline
No NER model N=10 & 0.628 & 9.11 \\
\hline
No NER model N=30 & 0.645 & 25.02\\
\hline
No NER model N=50 & 0.649 & 39.16\\
\hline

SpaCy \textit{sm} merged N=10 & 0.810 & 15.33\\
\hline
SpaCy \textit{sm} merged N=30 & 0.833 & 44.02\\
\hline
SpaCy \textit{sm} merged N=50 & 0.840  & 70.67\\
\hline
SpaCy \textit{sm} separate N=10 & 0.834 & 10.12\\
\hline
% SpaCy \textit{sm} separate + rel N=10 & 0.839  & 10.11 \\
% \hline

SpaCy \textit{trf} merged N=10 & 0.827 & 15.09\\
\hline
SpaCy \textit{trf} separate N=3 & 0.874 & 6.93\\
\hline
SpaCy \textit{trf} separate N=5 & 0.892& 11.68\\
\hline
SpaCy \textit{trf} separate N=10 & 0.911 & 23.47\\
\hline
% SpaCy \textit{trf} separate + rel N=10 & 0.913 & 23.44\\
% \hline

Flair merged N=10 & 0.861 & 15.54\\
\hline
Flair separate N=3 & 0.879& \textbf{6.27}\\
\hline
Flair separate N=5 & 0.895  & 10.58\\
\hline
Flair separate N=10 & \textbf{0.914}  & 21.30\\
\hline
% Flair separate + rel N=10 & \textbf{0.915}  & 21.28\\
% \hline

\end{tabular}%
}
\end{center}
\end{table}

Results of each configuration are described in Table~\ref{table5}. We decided to use the  "Flair ner-fast NER\_separate N=3" configuration. It provides high accuracy of 0.879 AR with only 6.27 candidates returned. While we are sacrificing 4\% of AR with respect to best results, the low N\_returned candidates are the most relevant value at this stage. Later in Table~\ref{table_ef}, we show that creating embeddings for candidates' articles is the most time-consuming piece of the system.

\subsection{The trade-off between Accuracy and Speed}

We need to use an NLI model that is fast, accurate, and can generalize well on an open-domain environment.  Here we want to understand the trade-off between accuracy and speed. We measured the accuracy and computation speed on a GPU (RTX2070) instance using the SNLI test set.

Initially, we reproduce the results of the NLI SOTA models and measure their efficiency. For that, we take SemBERT~\cite{ref_lncs19}, the top-performing word-based model, and the sentence-based HBMP model. 

Next, we compare SemBERT and HBMP with the proposed sentence-based Siamese architecture based on MLM (Section~\ref{model_two}). We use pre-trained MLMs coming from the Hugging Face platform\footnote{\url{https://huggingface.co/models}}.  The first one is \textit{bert-base-uncased}. This model is uncased, pretrained on the BookCorpus dataset~\cite{moviebook} and English Wikipedia~\cite{ref_lncs8}, this is a basic and lightweight model. The second one is the \textit{bart-base}, which is reported to work well for the summarizing tasks \cite{yoon2020learning}, which is related to the ability to interpret the text semantics, which can also be used on NLI problems.
Another model we tried is ALBERT, one of the top-performing models according to the GLUE score~\cite{lan2019albert}. It is also reported to be more memory efficient. We use "base" models as they are about three times faster than "large" ones for text encoding, according to our experiments.
We also include Universal Sentence Encoder (USE)~\cite{cer2018universal}, another popular sentence-based model.

\begin{table}[t]
\begin{center}
\caption{Experiments results on SNLI datasets.}
{\tabcolsep=3pt
\begin{tabular}{p{2.3cm}|p{1.8cm}|p{1.8cm}|p{1.8cm}}
\hline\label{table123}
\textbf{Model} & \textbf{SNLI dataset accuracy} & \textbf{CPU inference speed} & \textbf{GPU inference speed}\\
\hline
SemBERT & \textbf{91.9\%} & - & 0.510 $\frac{s}{sample}$ \\
\hline
HBMP & 86.6\% & - & 0.020 $\frac{s}{sample}$ \\
\hline
\textit{bert-base-uncased} & 85.2\% & 0.1 $\frac{s}{sample}$ & 0.006 $\frac{s}{sample}$\\
\hline
\textit{bart-base} & 86.9\% & 0.12 $\frac{s}{sample}$ & 0.006 $\frac{s}{sample}$\\
\hline
\textit{albert-base} & 84.98\% & 0.08 $\frac{s}{sample}$  & 0.006 $\frac{s}{sample}$\\
\hline
USE & 78.7\% & 0.036 $\frac{s}{sample}$ & \textbf{0.004} $\frac{s}{sample}$\\
\hline
\end{tabular}%
}
\end{center}
\end{table}
According to the results presented in Table~\ref{table123}, that although the word-based model SemBert is more accurate, the sentence-based models are 25x faster. Our Siamese architecture showed comparable accuracy and higher inference speed (efficiency), so we decided to proceed with the three of them for future experiments.  Interestingly, USE is the faster model, but the penalization on accuracy is high, so we decided to drop it. 

We also tested our three models' efficiency on  CPU (8G RAM, 4VCPUs). These results are important for our API, which is running on a small CPU instance on the Wikimedia Cloud\footnote{\url{https://wikitech.wikimedia.org/wiki/Portal:Cloud_VPS}}. We found that the \textit{albert-base} model is the fastest one.

\subsection{Model Generalization}

As we discussed in Section \ref{chap3}, datasets created to train and test NLI models have artifacts, hurting the ability of models to generalize in other scenarios. In order to understand the limitations of such datasets and design solutions to overcome this problem, here we study a Transfer Learning approach between the SNLI, MNLI, and FEVER datasets. To do this, we need to drop \textit{NOT ENOUGH INFO} class samples from FEVER because that category is not present on SNLI and MNLI. 

First, we trained a model (presented in Sec. \ref{model_two}) on the MNLI, tested on SNLI and MNLI testing set. We omit the details of those results due to lack of space. However, we found that the accuracy decays between 11\% to 16\% depending on the MLM used. Next, we tried training on SNLI and testing on the MNLI, FEVER, and SNLI. While the SNLI testing set results are around 85\% of accuracy for all language models, the same model tested on the MNLI test set gets around 60\%, a 20\% less than the model directly trained on MNLI. The results are even worse on the FEVER dataset, where the model trained on SNLI cannot go over 30\% accuracy. Models trained and tested on this 2-classes version of FEVER reach over 80\% accuracy.   

However, we want to highlight the BART models, which have the best generalization power. For example \textit{bart-base} model trained on SNLI has 86.9\% accuracy on SNLI and only 63.19\% on MNLI, which is about ~27\% drop, when we observe ~31\% drop for the \textit{bert-base-uncased} model.

As part of this experiment, we fine-tuned \textit{bert-base-uncased} and \textit{bart-base} on the WikiText dataset \cite{wikitext}. That is a collection of over 100 million tokens extracted from the set of verified high-quality articles on Wikipedia. Our goal here is to improve the performance of our models on Wikipedia's content.

The unsupervised fine-tuning was done on the RTX2070 GPU instance. To fine-tune, we followed the experiment setup by \cite{ref_lncs8} but trained only for the MLM problem formulation and just for one epoch. Reproducing the original setup, we selected 15\% of tokens at random. Then 80\% of tokens selected were changed to [MASK] special tag, 10\% were switched to another token, and 10\% remained original. 

\begin{comment}
Example of input and output such strategy is presented in Figure~\ref{fig:mlmt}.
\begin{figure}[!h]
\centering
\includegraphics[width=0.45\textwidth]{pictures/i13.png}\hfill
\caption{Example of input and output for masked sentence.}
\label{fig:mlmt}
\end{figure}
\end{comment}

As a result, we got two models fine-tuned for Wikipedia. As expected this tuning does not have an impact on non-Wikipedia related dataset such as SNLI (Table~\ref{tab:fulladap}), but improves   accuracy when they are tested on the FEVER dataset (see Tables~\ref{tab:fulladap}, \ref{tab:fevercln} and \ref{tab:feverfilt}).

One more experiment was to train the whole model on one dataset and then fine-tune (adapt) only the last dense layer (Figure~\ref{fig:figure_solution}), which is responsible for classification specifically for the target dataset. Using such a strategy is computationally efficient as the heavy MLM remains frozen but pre-trained on NLI tasks simultaneously. 

In our experiment, we train models on SNLI and then fine-tune the last layer on MNLI or FEVER train set and test on the corresponding test set. We also trained individual models for each dataset with all layers unfrozen and compared performance with the transfer learning approach. 
The results of this investigation are presented in Table~\ref{table44}.

\begin{table}[!b]
\begin{center}
\caption{Full training on specific dataset vs. training on SNLI and classifier adaptation on FEVER and MNLI}
\label{tab:fulladap}
{\tabcolsep=3pt
\begin{tabular}{|p{3.4cm}|p{2.2cm}|p{2.2cm}|}
\hline\label{table44}
\textbf{Model} & \textbf{MNLI adapted vs. full train} & \textbf{FEVER adapted vs. full train} \\
\hline
\textit{bert-base-uncased}  & 64.8\% / 76.1\% & 70.1\% / 79.81\%\\
\hline
\textit{bart-base}  & 67.6\% / 77.85\%& 74.4\% / 85.24\%\\
\hline
\textit{bert-base-uncased} + fine tuned  & 65.4\% / 76.29\% & 69.7\% / 82.45\%\\
\hline
\textit{bart-base} + fine tuned  & 68.1\% / 77.35\%  & 73\% / 85.62\%\\
\hline
\end{tabular}%
}
\end{center}
\end{table}

The transfer learning approach shows much lower results comparing to fully trained models on both MNLI and FEVER. It reveals that MLM plays a significant role in the whole NLI model and should also be adapted for every specific need, in our case, working with Wikipedia articles.

\subsection{Building Wikipedia NLI model}
As we showed in previous experiments, NLI models fully trained on a specific dataset perform much better than fine-tuned models. So we decided to train the FEVER-specific NLI model that will be the primary building block of a fact-checking system based on Wikipedia.

\subsubsection{Experiment setup}

For previous experiments, we used a model trained only for predicting two classes \textit{REFUTES (R)} and \textit{SUPPORTS (S)}, as there is no hypothesis presented for all samples of the \textit{NOT ENOUGH INFO (NEI)} class. Therefore, we need to generate samples for that class. We use a negative sampling strategy inspired by the approach followed by \textit{~\cite{fever3}}.

As for validation, we are using a predefined testing set. We are using another strategy for filling \textit{NEI} class in the testing set. We take the original claim for the \textit{NEI} samples, using model level one to pick article candidates for such sample and then randomly select one sentence from such articles. 

\subsubsection{FEVER Hypothesis cleaning}
The original sentences from the FEVER Wikipedia dump include tags at the end of the sentence. At the same time, the MediaWiki API does not return those tags along with texts, so we will not have them on our application. Therefore, we considered two types of training datasets: original hypothesis sentences and without tags (cleaned). Such cleaning decreased the average number of symbols in the hypothesis from 212 to 136 characters. As for validation scores, we also compare results on the cleaned and original test sets and separately evaluate performance on only $S$ and $R$ classes as they are not synthetic. 

From our experiment, we found out that cleaning tags from training data sets reduce the accuracy on test with tags and increase on the score without them. It means that training on cleaned texts is beneficial for the real-life solution. The results of training models on the cleaned dataset are presented in Table~\ref{table667}.

\begin{table}[b]
\begin{center}
\caption{Training on FEVER dataset with cleaned hypothesis}
\label{tab:fevercln}
{\tabcolsep=3pt
\begin{tabular}{|p{2.4cm}|p{1.2cm}|p{1cm}|p{1.4cm}|p{1.4cm}|}
\hline\label{table667}
\textbf{Model} & \textbf{FEVER original}  & \textbf{FEVER clean} & \textbf{FEVER original R\&S only}  & \textbf{FEVER clean R\&S only} \\
\hline
\textit{albert-base} & 72.40\% & 71.85\% & 67.11\% & 65.50\% \\
\hline
\textit{bert-base-uncased} & 71.97\% & 71.67\% & 67.17\% & 66.13\% \\
\hline
\textit{bart-base} & 74.20\% & 74.72\% & 68.11\% & 68.76\% \\
\hline
\textit{bert-base-uncased} + fine tuned & 72.02\% & 71.76\% & 67.01\% & 66.19\% \\
\hline
\textit{bart-base} + fine tuned & 74.18\% & \textbf{74.82\%} & 68.34\% & 69.33\% \\
\hline
\end{tabular}%
}
\end{center}
%\vspace{-24pt}
\end{table}

As we see, unsupervised fine-tuning of models using Wikipedia-specific text and hypothesis cleaning made a boost for \textit{bart-base} model. It is the best-performing model for a complete test and $R$ and $S$ classes only. 

\begin{figure}[!t]
\centering
\includegraphics[width=0.48\textwidth]{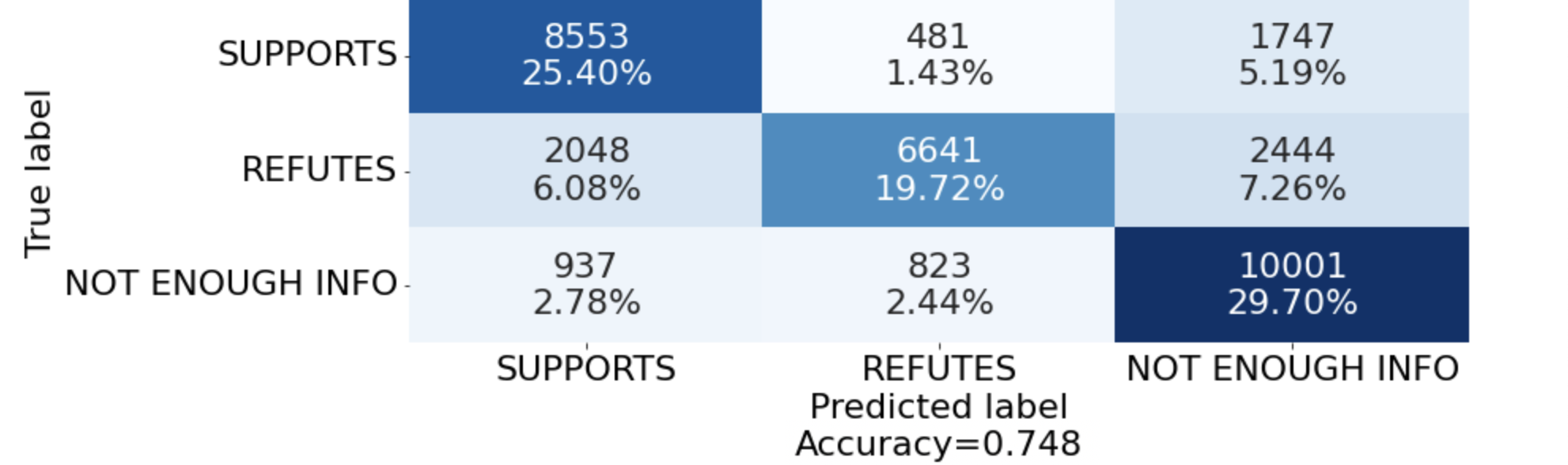}\hfill
\caption{Confusion matrix for \textit{bart-base} model (cleaned hypothesis training and validation).}
\label{fig:conf}
\end{figure}

We observed that the accuracy considering only the $R$ and $NEI$ classes is usually much lower than for the three classes. In order to understand the reasons behind that, we looked into the confusion matrix (Figure~\ref{fig:conf}). We found that the model has difficulties predicting the $R$ class as there are many false negatives. Approximately more than 13\% of accuracy we are losing just on the ($R$) class. We are solving this issue in the following experiment. It is essential to mention that models trained on the cleaned dataset generally perform much better when we consider only the $R$ and $S$ classes, which is another benefit of such pre-processing technique.

\subsubsection{FEVER filtering}
As it was shown in Figure~\ref{fig:figure3}, some hypotheses were repeated multiple times and correspond only to one class, which can lead to model over-fitting. As it was discussed by \cite{ref_lncs6}, such annotation artifacts have a significant impact on model accuracy. FEVER dataset has the same issue. So we considered filtering and balancing datasets and experiment with how it influences model performance. As a base dataset for filtering, we took the cleaned training dataset used in the previous section. Also, the same training procedure was done in order to get comparable results. 

We used three main steps during data filtering. Firstly, we filtered out absolute duplicates by fields "claim" and "hypothesis." That reduced number of samples by 8.8\% concerning the original size. 

After that, we proceed to filter samples with the duplicated hypothesis. For that, we selected the set of all samples with the same hypothesis sentence. Then we found the difference $N_{diff}$ in number of samples of \textit{SUPPORTS} and \textit{REFUTES} classes. Next, we picked the random number $N_{drop}$ from 0 to $N_{diff}$, which corresponds to the number of samples to drop. Then we randomly picked $N_{drop}$ samples to drop from major class in order to equalize the distribution of contradicting classes among one hypothesis. Such operation was done only for those hypotheses that correspond to at least ten samples. This reduces the dataset by an additional 6.9\%.% compared to the original FEVER size. 
\begin{figure}[!t]
\centering
\includegraphics[width=0.48\textwidth]{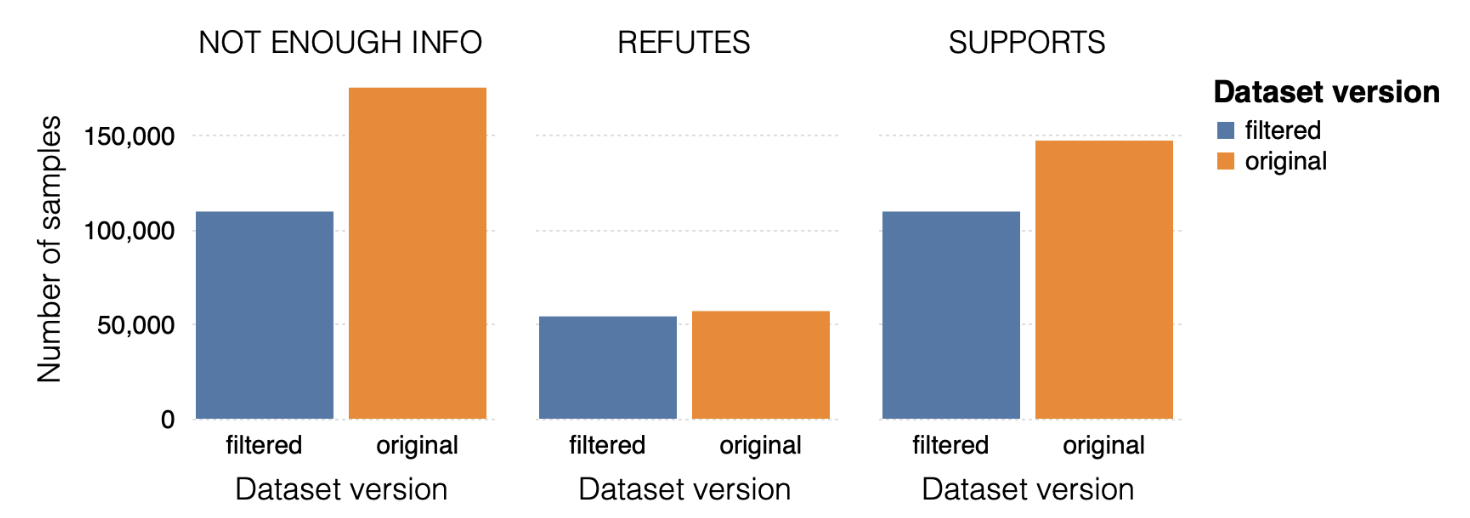}\hfill
\caption{Compare distribution of classes before and after filtering.}
\label{fig:filter_dis_compare}
\end{figure}
%\vspace{-24pts}

Finally, we randomly undersampled the number of \textit{NOT ENOUGH INFO} cases, equalizing it to the size of the \textit{SUPPORTS} class which is the second largest one. This action additionally decreased the filtered training dataset by 12.2\% with respect to the original size. As a result, we got the filtered dataset with the total number of samples reduced by 27.9\% concerning the original size.  

After that, we proceed with the training model on filtered data. We compared five main model results in Table~\ref{table_comparison}. We see that filtering significantly improves results for all configurations, especially for $R$ and $S$ classes. 

Finally, we decided to use the \textit{bart-base} + fine tune model, trained on cleaned, filtered FEVER dataset for WikiCheck API. This model showed the best accuracy on Wikipedia-related content, a reasonably good generalization power, and good time efficiency.

\begin{table}[!t]
\begin{center}
\caption{Training model on filtered FEVER vs. original (cleaned) dataset}
\label{tab:feverfilt}
{\tabcolsep=3pt
\begin{tabular}{|p{3cm}|p{2.2cm}|p{2.2cm}|}
\hline\label{table_comparison}
\textbf{Model} & \textbf{original vs. filtered} & \textbf{original vs. filtered R\&S only} \\
\hline
\textit{albert-base} & 71.85\% / 72.40\% & 67.11\% / 68.46\% \\
\hline
\textit{bert-base-uncased} & 71.67\% / 73.04\% & 67.17\% / 70.49\% \\
\hline
\textit{bart-base} & 74.72\% / 75.53\% & 68.11\% / 71.47\% \\
\hline
\textit{bert-base-uncased} + fine tuned & 71.76\% / 73.38\% & 67.01\% / 70.44\% \\
\hline
\textit{bart-base} + fine tuned & 74.82\% / \textbf{75.91\%} & 68.34\% / 71.91\% \\
\hline
\end{tabular}%
}
\end{center}
\end{table}

\subsection{WikiCheck: A complete Fact-Checking system based on Wikipedia}

In order to make our solution efficient in production environments, we considered:
\begin{enumerate}
    \item asynchronous processing of model level one in order to overcome I/O bound;
    \item using sentence-based models that allows calculating claim embedding ones only;
    \item batch processing for hypothesis embeddings calculation;
\end{enumerate}

Most of the SOTA solutions were created during the FEVER competition, where the main criterion for model comparison is accuracy. Our research orientation is shifted towards usability. It means that the model should be not only accurate but also fast and interpretable. 

All top solutions \cite{fever1,fever2,fever3} are using a three-staged model. These stages are article selection, sentence selection, NLI classification. 
We present a two-staged solution with ML-based aggregation on top. We consider using only document retrieval and NLI models for fact verification. We do not apply sentence selection logic to avoid missing important information and, what is more important, let the user decide by themselves.

Also, as for the NLI model, we are using a sentence-based approach. It significantly improves the speed of the NLI model on inference, sacrificing a little the accuracy of the results. Moreover, our simple NLI model achieves almost SOTA result for sentence-based models, being more efficient (Section~\ref{related:problem}). 

\subsubsection{Experiment setup}

There are two main characteristics of fact-checking applications that we want to measure: accuracy and time efficiency.

As for general fact-checking system accuracy validation, we used the original FEVER dataset. All experiments were done using the RTX2070 GPU instance. In order to measure application accuracy, we decided to use the official FEVER validation tool \footnote{ \url{https://github.com/sheffieldnlp/fever-scorer}}, which allows us to compare our solution with FEVER competitors. 

However, to make our results comparable with FEVER contest results, we need to add a step that selects the most relevant sentence for a given claim within the document and provide the judgment just on that evidence. To do this, we used CatBoost learning-to-rank model for evidence picking following the idea from \cite{chernyavskiy2021whatthewikifact}. This step is not included in our final API, and we use it only for comparison purposes. 

Also, as mentioned above, there are differences between the Wikipedia content in 2017 and the time of our experiments (May 2021), which is a handicap for our system. The same problem was reported by~\cite{fever3}. However, in our case, the time lag is more significant, and as a result, we have an 11.51\% of articles found by the MediaWiki API that do not have a matched text in the 2017 dump provided by FEVER.

It gives several metrics used for validation: 
\begin{enumerate}
    \item FEVER score: In order to consider the sample correctly classified, requires the correct label along with the full match of true evidence with predicted.
    \item Accuracy: Standard accuracy score that requires only label match to consider sample correctly classified.
    \item Evidence $F_1 @k$ score: The score that evaluates the correctness of picked evidence and does not take into account \textit{NEI} class samples. It calculates as following: $$F_1 @k = 2 \cdot \frac{(Precision @k) \cdot (Recall @k) }{(Precision @k) + (Recall @k)}$$ $$\text{Precision}@k = \frac{true \ positives \ @ k}{(true \ positives \ @ k) + (false \ positives \ @ k)}$$ $$\text{Recall}@k = \frac{true \ positives \ @ k}{(true \ positives \ @ k) + (false \ negatives \ @ k)}$$
\end{enumerate}

\begin{figure}[!t]
\centering
\includegraphics[width=0.47\textwidth]{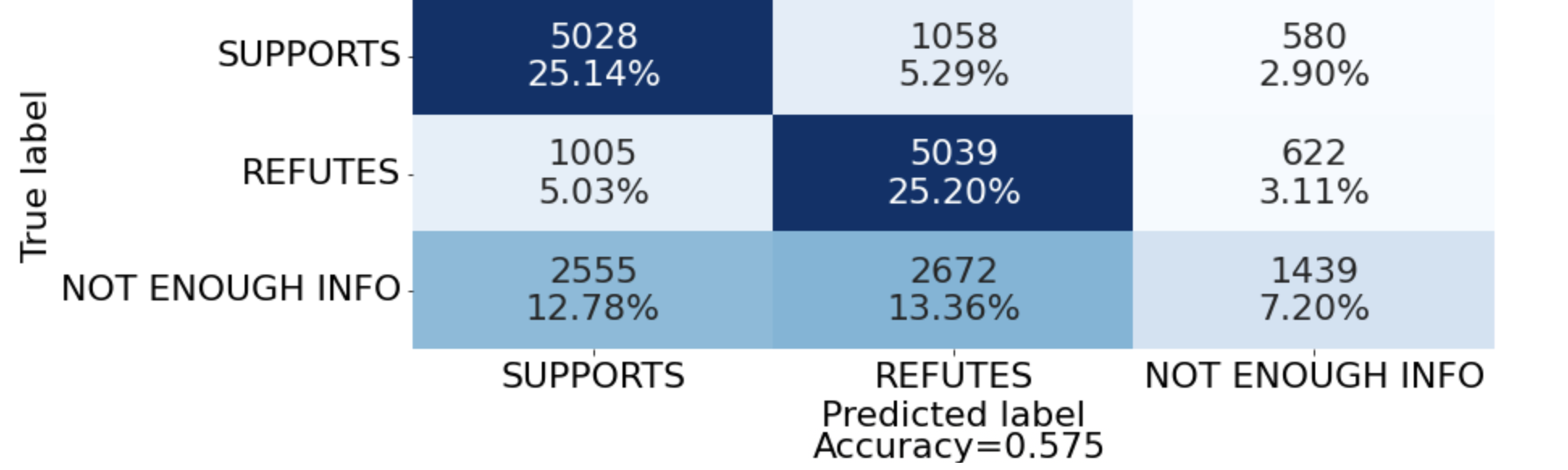}\hfill
\caption{Confusion matrix for \textit{WikiCheck} model}
\label{figure_conf}
\end{figure}

We compared our fact verification system (\textit{WikiCheck}) with the best performing solutions of FEVER competition. In comparison, we used our final model with a BART-based NLI classifier.  The final results are presented in Table~\ref{table_ac}. As a result, we got a 0.43 Fever score and 0.57 of general accuracy for our \textit{WikiCheck} model.  Also, we analyzed the errors of our models and found out that most of our mistakes are made for \textit{NEI} class. The confusion matrix for both of our models can be found in Figure~\ref{figure_conf}. We hypothesize this could be partially related to the differences between the Wikipedia dump in 2017 and current content. However, our results are comparable with top-8 FEVER results that are not focusing efficiency and generalization problems.  

\begin{table}[!t]
\begin{center}
\caption{Complete fact checking system FEVER accuracy}
\begin{tabular}{p{2cm}p{0.8cm}p{1cm}p{1cm}p{1cm}}
\toprule
Team/Name & FEVER rank &  Evidence F1 &  FEVER score &  Accuracy \\
\midrule
UNC-NLP &          1 &       0.5322 &       0.6398 &    0.6798 \\
UCL MRG &          2 &       0.3521 &       0.6234 &    0.6744 \\
Athene &          3 &       0.3733 &       0.6132 &    0.6522 \\
Ohio St. Uni &          7 &       0.5854 &       0.4322 &    0.4989 \\
\textbf{WikiCheck} &          - &       0.3587 &       0.4307 &    0.5753 \\
GESIS Cologne &          8 &       0.1981 &       0.4058 &    0.5395 \\
\bottomrule
\end{tabular}\label{table_ac}
\end{center}
\end{table}

We are also interested in measuring the speed of our system. We called "efficiency" the time (in seconds) used to perform a task.  We used a random set of one 1K unique claims from the FEVER test set. We tested the system using three configurations with different sentence encoder models. For each model, we used a different random set of claims to avoid API caching influence. The system was running on CPU-only 2,0 GHz Intel processor instance with 8Gb RAM provided hosted on the Wikimedia VPS Cloud\footnote{\url{https://wikitech.wikimedia.org/wiki/Portal:Cloud_VPS}}.

  We split the whole application into the logical part and tested each separately. \textit{NER\_model} part correspond to using the NER model for named entities extraction. MediaWiki API usage is represented by two parts: \textit{wiki\_search}, responsible for article search, and \textit{wiki\_text} corresponding to retrieving the texts for selected articles. Then we have two stages that represent embeddings calculation. We calculate embeddings for claim and hypothesis separately. The last step is \textit{classification} that in charge of using the NLI classifier given the sentence embeddings. 

The results for efficiency are described in Table~\ref{table_ef}.  The most time-consuming parts are retrieving the articles (\textit{wiki\_text}) and hypothesis embeddings calculation. Retrieving the articles' content takes about 40\% of total application time, and calculation embeddings for the hypothesis take 50\%.  The approximate time needed for the fact-checking process is about six seconds. Considering these results, for our API, we decided to use \textit{bart-base} model even though it was slighter slower, but it has the best accuracy.

\begin{comment}

    \begin{figure}[!t]
    \centering
    \includegraphics[width=1\linewidth]{pictures/i19.png}\hfill
    \caption{Fact Checking system efficiency}
    \label{fig:ef}
    \end{figure}
\end{comment}

\begin{table}[!t]
\begin{center}
\caption{Fact checking system efficiency (seconds)}
\begin{tabular}{l|rrr}
\toprule
        system parts  &    albert &      bart &      bert \\
\midrule
NER\_model &  $0.06\pm0.02$ &  $0.06\pm0.02$ &  $0.06\pm0.02$ \\
wiki\_search &  $0.39\pm0.20$&  $0.40\pm0.20$ &  $0.39\pm0.21$ \\
wiki\_texts & $2.40\pm1.14$ &  $2.37\pm1.06$ &  $2.30\pm1.10$ \\
embedding\_claim &   $0.07\pm0.01$ &  $0.08\pm0.02$ &  $0.07\pm0.01$ \\
embedding\_hypothesis  &  $3.05\pm1.36$ &  $3.18\pm1.59$ &  $2.56\pm1.28$ \\
classification &   $0.01\pm0.01$ & $0.01\pm0.01$ & $0.01\pm0.01$\\
           \midrule
total\_time & $5.97\pm2.35$ &  $6.11\pm2.47$ &  $5.41\pm2.24$ \\
\bottomrule
\end{tabular}\label{table_ef}
\end{center}
\end{table}

%\chapter{Conclusions and Future Work}

\section{Conclusions}
The main goal of this work was to transform academic research on NLI and Automated Fact-Checking into a usable automated fact verification tool, easy to use and does not require a lot of computational resources. Moreover, we focused on openness for all the components, including the knowledge base used as a source of ground truth. 

%We analyzed related research and defined open problems that should be solved to achieve our goal. 
Previous works have not put emphasis on the efficiency of their solution but concentrate on accuracy instead. The speed of the models is a crucial characteristic of practical application. Moreover, most SOTA solutions are implemented for GPU  processing  - which is expensive -  and require code refactoring to use CPU instances on inference. For end-users with limited resources, that make it impossible to use those solutions for their needs. The last but not the least problem is the scarcity of good NLI datasets to train models that work \textit{in the wild}. %The presented datasets have their specific limitations when creating a new dataset is expensive as it requires manual annotation. 

To overcome those problems, first, we performed an advanced data analysis to understand and try to solve the limitations of the most used datasets, adapting them to create a model that works in production environments. The main results of this work are:
%tried different modeling methods and discovered the following insights: 
\begin{itemize}
    \item We discovered that the FEVER dataset has annotation artifacts that can influence the model's performance. We proposed a filtering technique that increases the model's accuracy and generalization power.  
    \item  We designed a query enhancing technique that improves the evidence selection process and improves the time efficiency of the full system.  
    %We showed that NER models for search increase the quality of results, in case there are named entities in queries, as in the FEVER dataset. 
    \item We showed that SOTA NLI models have generalization problems. Therefore, we created an unsupervised fine-tuning heuristic that improves models' performance in real-life scenarios, especially working with Wikipedia. %Full model training is required to get the best results for a specific dataset. The only classification layer fine-tuning transfer learning does not work well. 
    %\item W
    %of MLM %ith domain-specific texts that further increased accuracy on the downstream NLI task.
    \item We found that the optimal compromise between time efficiency and accuracy is given by sentence-based language models and provides a set of heuristics to improve the efficiency of the full system. 
    %We showed that batch processing is faster than one-by-one. Also, we provided reasoning why sentence-based NLI models are faster than word-based for real-world applications. The proposed possible architecture of sentence-based NLI model that shows comparable to SOTA results, being more efficient.
    \item Finally, we present \textit{WikiCheck}, an end-to-end system for automated fact-checking using Wikipedia as the knowledge base. Furthermore, we designed a NER-based solution that improves evidence discovery on the MediaWiki Search API.
\end{itemize}

 %that can receive a sentence (claim), query Wikipedia is looking for evidence, and then apply the NLI model on that pair (\textit{claim hypothesis}), returning the relation of each corresponding pair, which is one of \textit{SUPPORT}, \textit{REFUTED} or  \textit{NOT ENOUGH INFO}. 
Our solution has comparable SOTA results. It can be used on CPU, low memory devices, which makes it more applied. We make all the code for WikiCheck API available on Github. We provided a detailed README that will allow us to reuse our code easily. Also, we make our system used as an open API. %\footnote{WikiCheck Github repository \url{https://github.com/trokhymovych/WikiCheck}.}.

%However, the system has relevant limitations. The system should be faster. % We should research more on the aggregation stage. 
%We have problems with $NEI$ labeled samples, which is a limitation for real-life scenarios. More research should be done for this part of the application. Moreover, WikiCheck API depends on the MediaWiki Search API that we do not have control over.

Nonetheless, we acknowledge important limitations in our work. First of all, we are heavily dependent on Wikipedia, both for content and search. Moreover, although we focus on efficiency, the average of 6 seconds per claim verification should be improved in future research. Finally, the SOTA performance of NLI models is still failing for around 33\% of the cases, making it difficult to rely on these systems in production environments completely. However, we consider that by creating this open API, we generate awareness of the opportunities, risks, and limitations of using the automated fact-checking system in real life.

\bibliographystyle{ACM-Reference-Format}
\bibliography{master-thesis-template-blx}
\end{document}